\documentclass[global,twocolumn]{svjour}
\usepackage{latexsym}
\usepackage{graphics}

\usepackage{amsmath}
\usepackage{amssymb}
\usepackage[latin1]{inputenc}
\usepackage{tikz}
\usepackage{graphicx}
\usepackage{bm}      
\usepackage{color}
\usepackage{soul}
\usepackage{hyperref}
\usepackage[numbers]{natbib}
\usepackage{calc}    
\usepackage{textcomp}
\usepackage{breakcites}
\usepackage{relsize} 
\journalname{}
\begin{document}
\title{Precision spectroscopy technique for dipole allowed transitions in laser cooled ions}
\author{Amy Gardner \and Kevin Sheridan \and William Groom \and Nicolas Seymour-Smith \and Matthias Keller}
%
%
\institute{ITCM Group, Department of Physics and Astronomy, University of Sussex, Falmer, BN1 9QH, England.}
\date{Received: date / Revised version: date}
%
\maketitle
\begin{abstract}
In this paper we present a technique for the precise measurement of electric dipole allowed transitions in trapped ions. By applying a probe and a cooling laser in quick succession, the full transition can be probed without causing distortion from heating the ion. In addition, two probes can be utilized to measure a dispersion-like signal, which is well suited to stabilizing the laser to the transition. We have fully characterised the parameters for the measurement and find that it is possible to measure the line center to better than 100~kHz with an interrogation time of 30~s. The long-term stability of the spectroscopy signal is determined by employing two independent ion trap systems. The first ion trap is used to stabilize the spectroscopy laser. The second ion trap is then employed to measure the stability by continuously probing the transition at two frequencies. From the Allan variance a frequency instability of better than 10$^{-10}$ is obtained for an interrogation time of 1000~s.
\end{abstract}
\section{Introduction}
\label{intro}
The precise characterization of electric dipole allowed transitions of atomic ions is an indispensable tool in modern ion trap based technologies such as quantum information processing \cite{Turchette,Haffner}, quantum optics \cite{Bloch} or collision studies \cite{Hall}. Measurements of the transition line shape allow for the precise determination of the central frequency of the spectral line and the optimal detuning of the lasers used for Doppler cooling.  In addition, the Lorentzian and Gaussian contributions to the line shape can be utilized to determine the magnitude of laser intensity broadening and measure the thermal motion respectively. Furthermore, measurements of the fluorescence spectrum have been used to accurately determine the heating rates of ions in traps \cite{Wesenberg,Brama}. 

While high resolution spectroscopy on forbidden transitions has been very successful \cite{Roos,Rosenband,Margolis,Stenger}, the determination of the transition wavelength of strong transitions is challenging. Due to the large scattering rates of these transitions, the spectroscopy laser not only probes the atomic transition but also alters the motional state of the trapped ions due to recoil heating when particular parts of the spectrum are probed. This in turn changes the spectral line shape and thus hampers the precise measurement of the transition frequency. There have been a few different approaches to this problem, for instance Wan \textit{et al.} have recently devised a scheme to transfer the scattering rate of a spectroscopy ion to a co-trapped read-out ion to measure the full, undisturbed spectrum of a single $^{40}$Ca-ion \cite{Wan}. Other approaches include using single-photon scattering \cite{Hempel,Clos}, using separate probe and cooling lasers \cite{Herrmann,Drullinger}, or scanning over the resonance to produce a half-Lorenztian lineshape \cite{Nagourney}. 

In this paper, we present a technique to measure the spectrum of trapped ions without significant heating of the ion, and thus avoiding line shape distortion. We show that it is possible to measure the line center to better than 100~kHz with an interrogation time of 30~s. Employing two independent ion trap systems, we demonstrate that a frequency instability of less than $10^{-10}$ within 1000~s can be achieved.

In the first section of this paper we present the measurement principle, which is followed by a description of the experimental set-up. Finally, the characterization of the spectroscopy method is described in Sec. \ref{section4}.

\section{\label{sec:measurement_principle}Measurement principle}

The standard means of measuring the fluorescence spectrum of strong transitions is to scan the frequency of the spectroscopy laser over the resonance while recording the ion's fluorescence signal.  Cooling the ion while simultaneously collecting a fluorescence spectrum is challenging since scanning the frequency of the spectroscopy laser changes the cooling dynamics. As the cooling laser frequency is scanned across the transition from red to blue detuning, the Doppler cooling efficiency peaks, rapidly decreases, and eventually leads to heating. The decrease in cooling efficiency leads to an increase in the temperature of the ion. Near the line center the rate of heating surpasses the cooling rate, resulting in a rapid loss of ion localization \cite{Waki}. This is accompanied by a loss of the fluorescence signal.  A Lorentzian function may be fit to the measured fluorescence spectrum to determine the linewidth and line center. However, due to the restricted range of the measured fluorescence spectrum, the line centre and linewidth can only be determined with limited precision. As a further drawback, the ion may gain enough energy through heating during the measurement to escape from the trap. This is a particularly important consideration when shallow trapping potentials are employed.

The complications arising from the standard spectrosc- opy procedure described above can be removed by modifying the experimental set-up to separate the laser cooling and spectroscopy. One laser beam is employed to probe the transition while the frequency of the other laser beam is maintained at an optimal detuning for Doppler cooling. 

We have developed a spectroscopy technique similar to the method of Wolf \textit{et al.} \cite{Wolf} that requires only a single laser beam, which serves both as the Doppler cooling laser and as the probe. The laser frequency is rapidly switched between cooling and probing by employing a double-pass acousto-optical modulator (AOM) set-up, controlled by a pair of voltage controlled oscillators (VCOs).  The entire fluorescence spectrum of a single trapped calcium ion can be measured with high precision while maintaining a low ion temperature and good ion localization throughout the scan. Furthermore, the laser parameters can be independently adjusted to optimize the cooling and also the spectroscopy signal.
 
This scheme can be extended to include a second probe interval, during which the frequency is slightly detuned from the first probe frequency. Thus the ion's fluorescence is probed at two laser detunings in quick succession by consecutive probe pulses interrupted only by a short cooling interval.
Subtracting the fluorescence rates during both probe intervals results in a dispersion-like signal for the ion's transition. This scheme is robust against drifts of the laser power and directly provides an error signal to precisely determine the line centre of the transition and stabilize the probe laser.


\section{\label{sec:exp_set_up}Experimental set-up}
\begin{figure}[ht]
\begin{center}
\resizebox{0.50\textwidth}{!}{\includegraphics{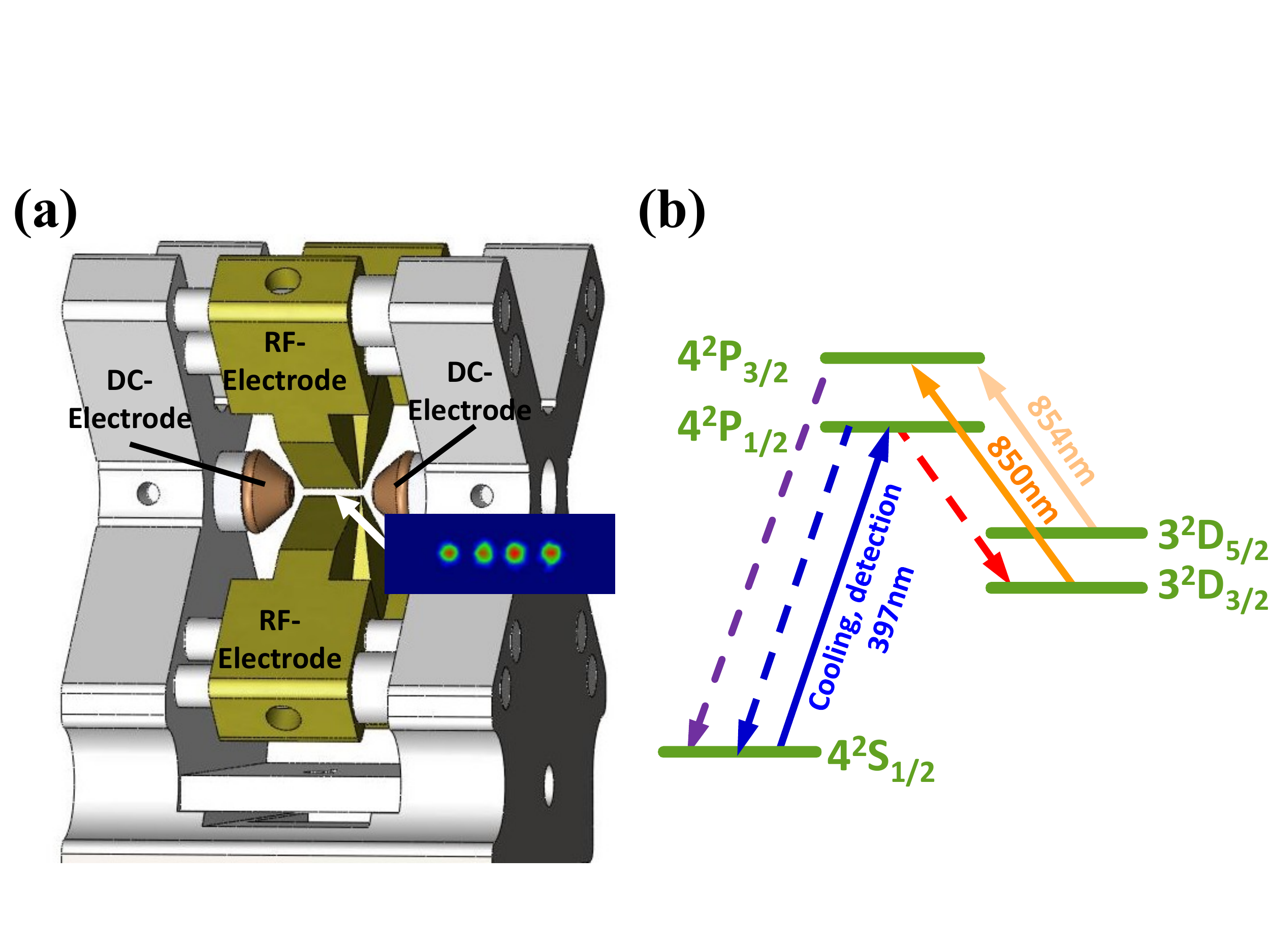}}
\caption[]{(a) Schematic of the ion trap. (b) Level scheme of $^{40}$Ca-ion including all lasers required to cool and re-pump the ion (solid arrows). Dashed lines indicate spontaneous emission.}
\label{fig:iontrap}
\end{center}
\end{figure}

The trap used in this experiment is a linear rf Paul trap. A schematic of the ion trap is shown in Fig. \ref{fig:iontrap}(a). It consists of four blade shaped rf-electrodes, which provide the radial confinement. The ion-electrode separation is 465~$\mu$m and the rf-electrode length is 4~mm. Positive static potentials applied to two DC-electrodes are used for the axial confinement. Through each of the DC-electrodes is an aperture that provides laser access to the trapped ions along the trap axis.

The trap is operated at a frequency of 26~MHz and a typical rf-power of 300~mW to obtain a radial secular frequency of 1.5~MHz. The typical axial confinement is 250~kHz at a DC-voltage of 200~V. For appropriate trapping parameters, single ions, linear ion strings and large three-dimensional Coulomb crystals with several hundred ions can be quickly and reliably loaded into the trap.

\begin{figure}[ht] 
\begin{center}
\resizebox{0.45\textwidth}{!}{\includegraphics{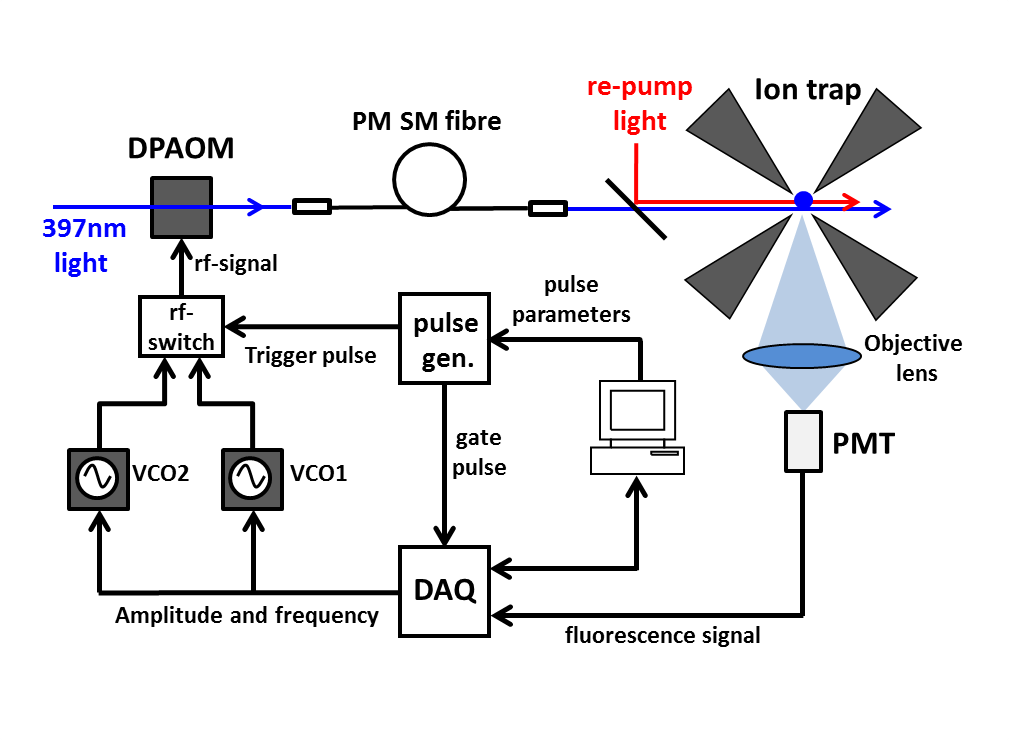}}
\caption{Schematic of the experimental set-up. The 397~nm light is coupled through a double-pass AOM (DPAOM) set-up before it is sent to the ion trap through a single-mode (SM) polarization maintaining (PM) optical fiber. The rf-signal is supplied to the AOM by either of two VCOs connected through an rf-switch. A computer controlled pulse generator switches between the two VCOs by supplying a trigger pulse to the rf-switch. A gate pulse, synchronous with the trigger pulse, is supplied to the DAQ to ensure that the fluorescence signal from the PMT is only recorded when the probe VCO is active.}
\label{figS:spec_setup} 
\end{center}
\end{figure}

Loading of the trap occurs through the photoionisation of neutral calcium atoms effusing from a resistively heated oven located below the trap center and aligned perpendicular to the trap axis. Photoionization is a two-photon process. The first photon is resonant with the S$_{0}\rightarrow$ P$_{1}$ transition of neutral calcium at 423~nm and the second photon at 375~nm provides the energy necessary to ionize the excited state calcium atoms \cite{Lucas}. 

The $^{40}$Ca-ion is Doppler-cooled on the 4S$_{1/2}\rightarrow$ 4P$_{1/2}$ transition at 397~nm (Fig. \ref{fig:iontrap}(b)) by two laser beams, one aligned radially with a beam waist of 50~$\mu$m  and one aligned along the trap axis  with a beam waist of 70~$\mu$m. Consequently, all directions of the ion's motion are cooled. In order to avoid populating the meta-stable D$_{3/2}$-state through the decay of the P$_{1/2}$-level, a re-pump laser on the D$_{3/2}\rightarrow$ P$_{3/2}$ transition at 850~nm is applied. From here, the ion is returned to the ground state through the spontaneous decay of the P$_{3/2}$-state. However, another re-pump laser at 854~nm must be applied in order to avoid optical pumping into the meta-stable D$_{5/2}$-state. This re-pump scheme has the advantage over re-pumping via the P$_{1/2}$-level in that it creates an effective two-level system without coherence effects due to the coupling of the cooling and re-pump transitions.

The cooling and probe laser set-up is shown in Fig. \ref{figS:spec_setup}. The frequency and amplitude of the 397~nm laser is controlled by an AOM in double-pass configuration and is fiber coupled to the ion trap. The fiber output is split and aligned along the axis and  in the radial plane of the trap. The laser cools a single calcium ion in the center of the trap. The ion's fluorescence at 397~nm is collected by an objective lens with a numerical aperture of 0.22. The total detection efficiency of the set-up is about 0.1\%. Photons are counted with a photomultiplier tube (PMT). The PMT signal is recorded with a data acquisition device (DAQ)\footnote{National Instruments USB-6229} and the timing of the measurement is controlled by a pulse generator\footnote{SRS-DG5920 four-channel pulse generator}.

\begin{figure}[ht]
\begin{center}
\resizebox{0.45\textwidth}{!}{\includegraphics{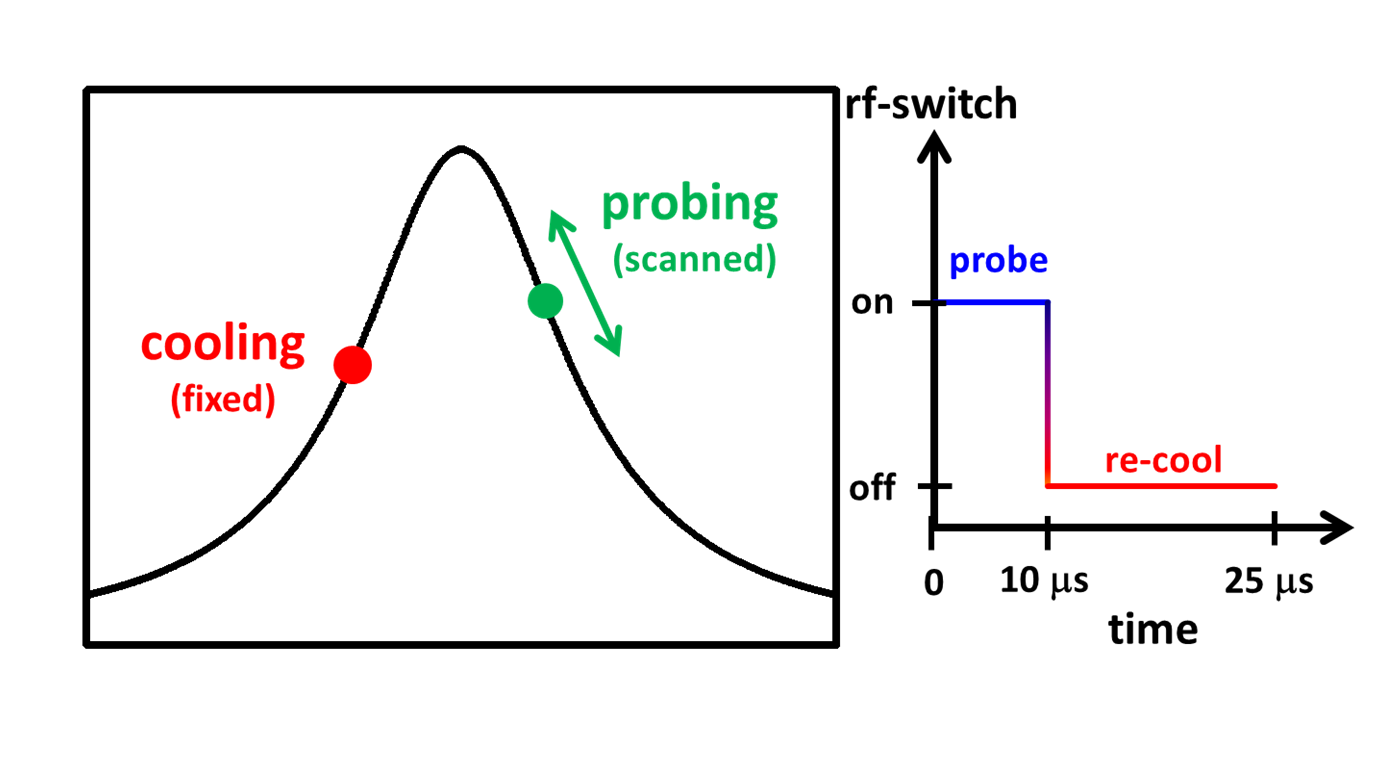}}
\caption[]{\textit{Left}: The cooling VCO maintains a frequency shift selected for optimal Doppler cooling and remains fixed for the duration of the scan. The probe VCO is stepped over a frequency range covering the entire fluorescence spectrum. \textit{Right}: A typical pulse sequence used during a spectroscopy measurement. The AOM is controlled solely by the probe VCO when the trigger pulse supplied to the rf-switch is on. When the pulse supplied to the rf-switch is off, control of the AOM is switched back to the cooling VCO and the ion is re-cooled.}
\end{center}
\label{figS:spec_pulse_sequence}      
\end{figure}

Two VCOs are connected through an rf-switch\footnote{Mini-Circuits ZASWA-2-50DR DC-5GHz} to the AOM. The rf-signal amplitude and frequency of each VCO output is controlled by analogue signals from the DAQ. The cooling VCO maintains a constant frequency that is selected for optimal laser cooling. The frequency output of the probe VCO is stepped over the probe frequency range, which typically spans 160~MHz. A square pulse from the pulse generator is applied to the rf-switch to switch control of the AOM from the cooling VCO to the probe VCO for the duration of the pulse. An identical pulse is simultaneously applied to the DAQ and acts as a gate for the PMT signal. Therefore, photon counts from the PMT are only recorded during the probe pulse. Following a probe pulse, control of the AOM is switched back to the cooling VCO and the ion is re-cooled. 

The 397~nm laser power must be kept the same for each probe frequency step in the scan to avoid distortion of the spectrum. However, the laser power in the beam diffracted by the AOM varies as a function of the applied frequency shift.  In order to produce a constant 397~nm laser power during the frequency scan, the amplitude of the rf-signal input to the AOM is scanned concurrently with the frequency. 
During calibration, the frequency is stepped while simultaneously measuring the 397~nm laser power.  An rf-amplitude value giving constant laser power is determined for each step in the frequency scanning range. The laser power remains constant with fluctuations smaller than 2\% of the mean. 

\begin{figure}[ht]
\centering
 \includegraphics[width= 1\hsize,angle=0]{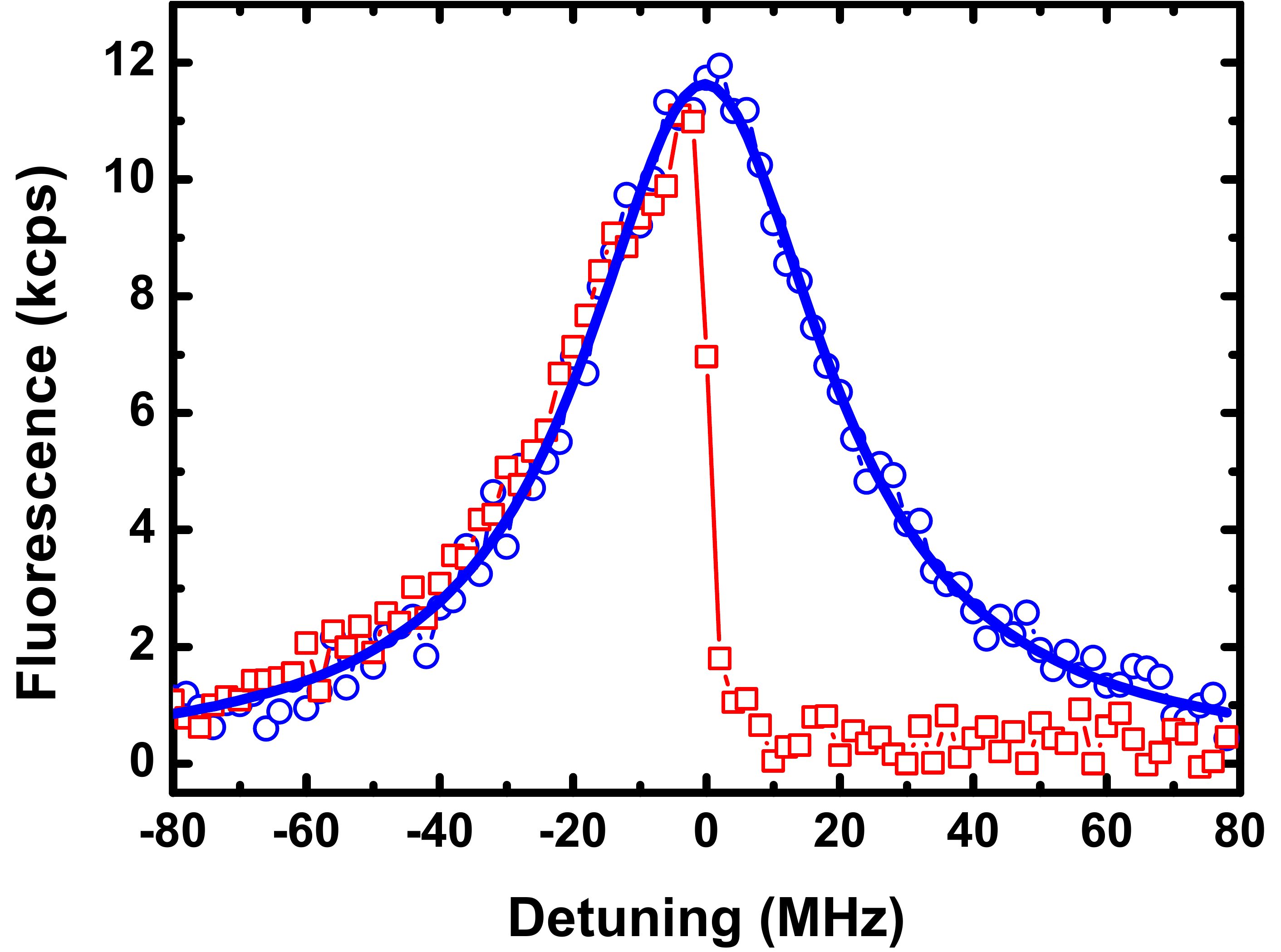}
\caption{Spectrum of the S$_{1/2}\rightarrow$ P$_{1/2}$ transition in calcium. The spectrum obtained by scanning the spectroscopy laser without additional cooling intervals ({\color{red}-}${\color{red}\Box}${\color{red}-}) shows a sharp drop close to the line centre. Employing additional cooling intervals results in a full, undistorted spectral line ({\color{blue}-}${\color{blue}\bigcirc}${\color{blue}-}). A Voigt profile is fit to the undistorted spectrum (blue line).}
\label{fig:spectroscopy_example}
\end{figure}

\begin{figure}[ht]
\centering
 \includegraphics[width= 1\hsize,angle=0]{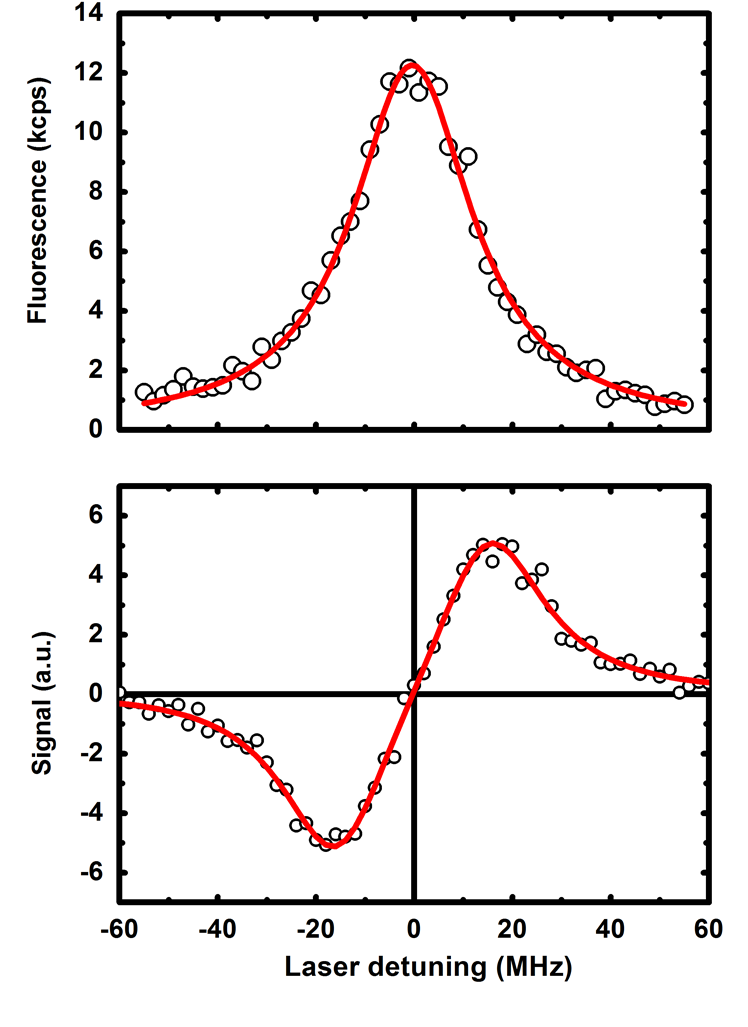}
\caption{Spectrum of the S$_{1/2}\rightarrow$ P$_{1/2}$ transition in calcium (top). Over the same range, two probes are used to obtain a dispersion-like signal (bottom). For this 5~MHz resolution scan, the two probes are set at 30~MHz apart. Subtracting the fluorescence count for one of the probes from the other generates the dispersion-like signal.}
\label{figS:spectroscopy_dispersion}
\end{figure}

Figure \ref{figS:spec_pulse_sequence} shows a schematic of the pulse sequence. Typically, the pulse sequence is repeated at a rate between 5~kHz and 40~kHz. On the order of 10,000 probe pulses of duration between 5~$\mu$s and 25~$\mu$s are applied for each probe frequency and the photon counts for all pulses are summed to produce a single data point in the fluorescence spectrum. After collecting a data point, the probe frequency is stepped and the process is repeated until the entire fluorescence profile has been measured. Finally, a Lorentzian is fit to the collected fluorescence spectrum and the line center and linewidth are recorded.

Figure \ref{fig:spectroscopy_example} shows an example of a 2~MHz resolution fluorescence spectrum. A single calcium ion is probed over a 160~MHz range during a scan with a total run-time of 40~s. For each step of the scan, 25~$\mu$s of probing is followed by 175~$\mu$s of re-cooling. A Voigt profile has been fit to the data and the measured power-broadened width of the spectrum is 44.5(5)~MHz. The natural linewidth of this transition is 21.6~MHz \cite{Wan}. In contrast to the spectroscopy scan without a cooling interval (red curve), the spectrum with intermediate cooling intervals has a Lorentzian line shape. The Gaussian contribution of the Voigt profile is much smaller than the fit error and is thus negligible.

A third VCO is added to the set-up to measure a dispersion-like signal. The same cooling/probing sequence is used but now the probe laser is switched between two frequencies by an additional rf-switch. Two independent fluorescence counters are used to measure the fluorescence rate during both probe intervals separately and the resulting rates are subtracted.  The parameters of the sequence are the same as for the previous measurement. After collecting a data point, the probe frequencies are changed synchronously and the process is repeated until the entire fluorescence profile has been measured. Figure \ref{figS:spectroscopy_dispersion} shows a 5~MHz resolution spectrum. The top graph shows the fluorescence spectrum measured simultaneously with the dispersion-like signal. The bottom graph shows the dispersion-like signal with its characteristic zero crossing at the resonance frequency. A single calcium ion is probed over a 120~MHz range by a scan with a total run-time of 40~s with a laser power of 6(2)~$\mu$W in each laser beam.
The measured Lorentzian linewidth of 29.6~MHz is in good agreement with the expected saturation broadening. Other contributions to the line broadening are expected to be significantly smaller than the power broadening. A detailed discussion of line broadening and systematic shifts is presented in Sec. \ref{section4.2}.
By probing two parts of the transition simultaneously in this manner, it is possible to measure the position of the line center constantly over time. This method is robust against fluctuations in laser power and can be used to stabilize the frequency of the 397~nm laser (Sec. \ref{section4.2}).

\section{Measurement characterisation}
\label{section4}

\subsection{Pulse sequence}

The most important parameters to consider when optimizing the spectroscopy measurement are the duration and repetition rate of the probing. For each step in the frequency scan a sufficient fluorescence signal must be collected while maintaining a cool and well localized ion. For probe frequencies close to line center or on the blue side of the line center, the ion's temperature will increase due to momentum kicks from photon scattering events and the lack of laser cooling. In order to maintain a low ion temperature, an upper limit on the duration of each probe pulse must be set.

Each probe pulse is followed by a re-cooling pulse that removes the measurement-induced motion of the ion. The re-cooling pulse must be long enough to fully restore the ion's equilibrium state. Otherwise, an accumulation of thermal energy throughout the course of the scan will occur. To avoid this `pile-up' of thermal energy during a scan, a lower limit on the ratio of the re-cooling time to probing time must be established. Additionally the probe pulse length must be sufficiently short to reduce the heating of the ion during blue detuning.

For a pulse sequence with sufficient re-cooling time, the spectral line shape is Lorentzian with a width determined by contributions from the natural linewidth of the transition and homogeneous broadening. Inhomogeneous contributions to the line shape, caused by the thermal motion of the ion, are not detectable in the fluorescence spectrum if the ion temperature remains low. Thermal motion of the ion, due to laser heating through long probing durations, will only occur for probe frequencies close to line center and on the blue side of the resonance peak. Therefore, for the case of excess thermal motion, inhomogeneous broadening will only contribute to the line shape in this region. The result is a fluorescence spectrum with a non-Lorentzian, asymmetric shape. 

We employ the coefficient of determination (R$^{2}$) of the Lorentzian fit to detect the presence of distortion in the spectral line shape due to heating effects. For a spectrum with N measured fluorescence values $y_i(x_i)$  at frequency $x_i$ the coefficient of determination is defined as:
\begin{equation}
R^2=1-\frac{\sum\limits_{i=0}^N \left\{y_i(x_i)-f(x_i)\right\}^2}{\sum\limits_{i=0}^N \left\{ y_i(x_i)-\bar{y}\right\} }
\end{equation}
where the fitting function is $f(x_i)$ and $\bar{y}$ is the mean value of the data points. As the fit residuals approach zero, R$^{2}$ approaches 1.

The fluorescence spectra for a range of pulse sequence parameters were collected. The results are used to determine both an upper limit on the probe time and also the ratio of probe to re-cooling time. 

\begin{figure}[ht]
\begin{center}
\resizebox{0.49\textwidth}{!}{\includegraphics{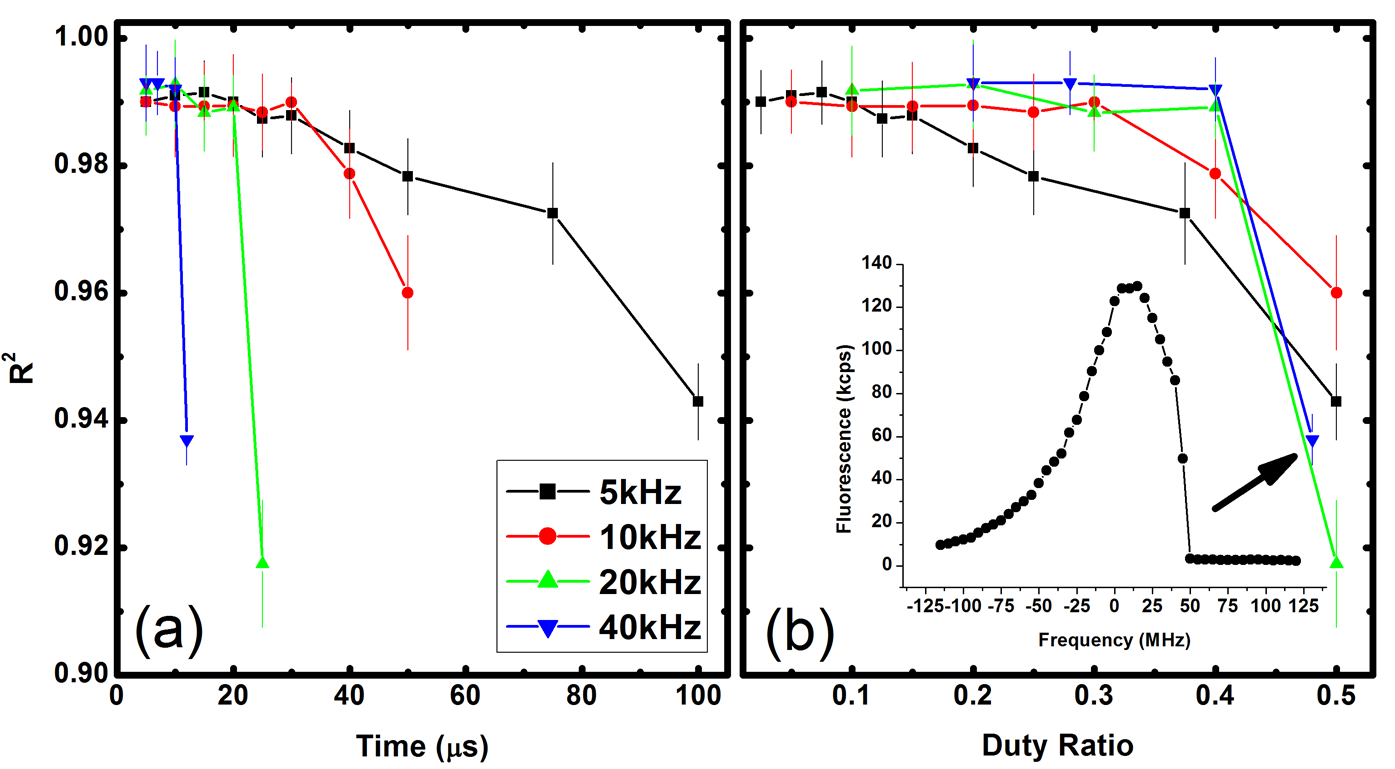}}
\end{center}
\caption[]{(a) R$^{2}$ as a function of probe time for a range of pulse sequence repetition rates. For repetition rates of 20~kHz and 40~kHz the value of R$^{2}$ drops sharply at small probe times. For repetition rates of 5~kHz and 10~kHz R$^{2}$ begins to decrease when the probe time exceeds 35(5)~$\mu$s. (b) R$^{2}$ as a function of duty ratio. As the repetition rate is reduced, a smaller duty ratio is required to sufficiently re-cool the ion. \textit{Inset:} an example of distortion of the line shape.}
\label{figS:spectroscopy_fit_distortion}      
\end{figure}

Figure \ref{figS:spectroscopy_fit_distortion}(a) shows the R$^{2}$ value of the Lorentzian line shape fit to the measured fluorescence spectrum as a function of the probe time for a range of duty ratios (probe time/period). Part (b) of the figure shows the R$^{2}$ value as a function of the duty ratios used in part (a). Each data point is the mean R$^{2}$ value for ten fluorescence spectra of a single ion. The error bars are the standard deviation of each data set. A single 397~nm laser aligned in the radial plane of the trap with a constant laser power of 6(2)~$\mu$W is used for each measurement. A fixed laser detuning of $-15$~MHz is used during the cooling portion of each pulse sequence. A frequency range covering 160~MHz is probed in steps of 2~MHz for each spectrum. 

For pulse sequence repetition rates of 20~kHz and 40~kHz the value of R$^{2}$ is consistently greater than 0.99(1) until it drops sharply for probe times of 25~$\mu$s and 10~$\mu$s respectively, equivalent to a duty ratio of 0.4. This drop is caused by a distortion of the fluorescence spectrum due to the heating pile-up effect discussed above. When the probe frequency is tuned to the point on the spectrum corresponding to maximum heating, the ion's equilibrium state can only be fully restored with a re-cooling detuning selected for maximum cooling, as is the case for these measurements. In addition, the duration of cooling must be at least equal in duration to that of the probing. 

For repetition rates of 5~kHz and 10~kHz, R$^{2}$ decreases at a slower rate with increasing duty ratio, compared to the sharp drop-off seen for 20~kHz and 40~kHz. The drop-off starts much earlier for the slower repetition rates, despite the fact that the cooling time is longer than the probing time. Even though the ion may be fully restored to the equilibrium state following probing, for long probe durations heating of the ion leads to a significant Gaussian contribution to the total line shape. As a result, the fluorescence rate during probing will change as a function of time due to the increased width and decreased amplitude of the fluorescence spectrum. Since this effect only occurs for probe frequencies near line center or on the blue side of line center the resulting measured fluorescence spectrum will have a distorted shape. 

The data in Fig. \ref{figS:spectroscopy_fit_distortion}(a) shows that this effect becomes measurable as a drop in R$^{2}$ for probe durations between 30~$\mu$s and 40~$\mu$s. This sets an upper limit of 35(5)~$\mu$s on the duration of probing, which gives duty ratios of 0.2 for 5~kHz repetition, and 0.4 for a repetition of 10~kHz. For probe durations beyond 35(5)~$\mu$s the measured spectrum will be distorted regardless of the re-cooling duration due to the temperature increase of the ion during probing. This can clearly be seen in Fig. \ref{figS:spectroscopy_fit_distortion}(b). As the repetition rate is lowered, a smaller duty ratio is required to sufficiently re-cool the ion.

From numeric simulations of the system at short probe times (\textless 30~$\mu$s) and low duty ratios, an average temperature increase of the ions of less than 2~mK is expected. Specifically, for a 25~$\mu$s long blue detuned probe pulse at low intensity (saturation parameter s = 0.1) the average temperature during the probe pulse increases by less than 1~mK. This is consistent with our previous findings \cite{Sheridan2}.

\subsection{Stability}
\label{section4.2}

In order to determine the long-term stability of the spectroscopy signal, we employ two independent ion trap systems. The first system is employed to generate a dispersion-like signal by measuring the transition at two frequencies equally detuned either side of line center. This signal is used to stabilize the spectroscopy laser. To maintain low ion temperature, the red detuned laser is applied for 175~$\mu$s while the blue probe laser is applied only for 25~$\mu$s. This sequence is repeated with a frequency of 5 kHz. By feeding back the error signal to the laser through digital PID control, the laser is stabilized to the atomic resonance. The feedback bandwidth is a compromise between the statistical fluctuation of the measured count rate and the drift of the free running laser. We have determined the optimal interrogation time to be 300~ms. The laser is stabilized to the line center of the transition to mitigate the effect of power broadening on the signal.

To measure the absolute frequency stability of the system we performed a long-term spectroscopic measurement in an independent ion trap with the stabilized 397~nm laser. The 4S$_{1/2}\rightarrow$ 4P$_{1/2}$ transition is probed by the same method as used in the first trap. The dispersion-like error signal generated is employed to measure any changes in the laser detuning with respect to the ions resonance. The two identical traps are about 5~m apart on separate benches and both systems have independent control systems for the spectroscopy laser beams.

\begin{figure}[ht]
\begin{center}
\resizebox{0.5\textwidth}{!}{\includegraphics{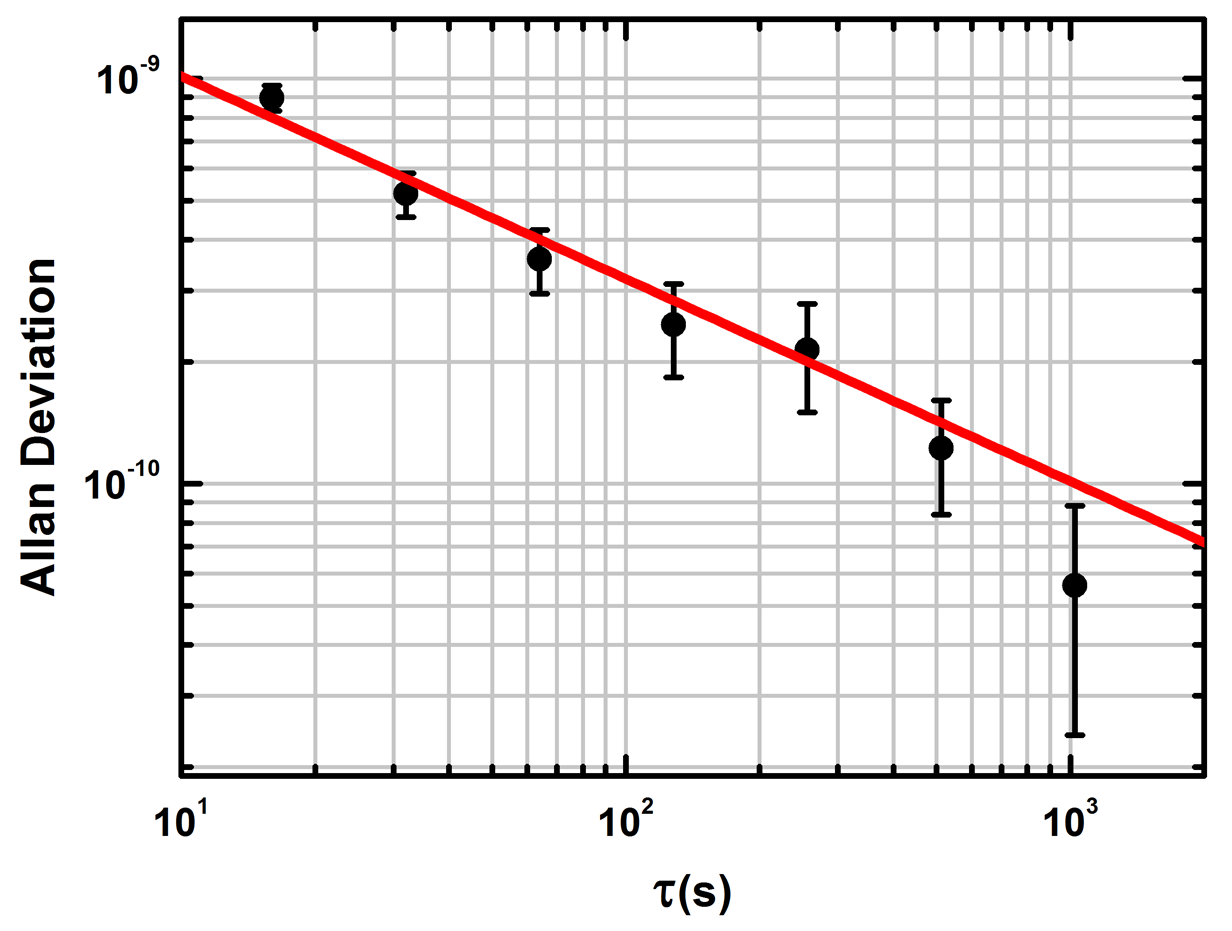}}
\caption{Measurement of the Allan variance of the spectroscopy signal of a single trapped calcium ion.}
\label{fig:independant_allan_variance}
\end{center}
\end{figure}

Figure \ref{fig:independant_allan_variance} shows the Allan variance for the spectroscopy signal. From a fit we obtain an instability of our spectroscopy signal of $\sigma(\tau)=3.5\cdot 10^{-9}\sqrt{s/\tau}$, and thus a statistical frequency uncertainty of $6\cdot10^{-11}$ within 1000~s. This is in keeping with the recent work by Wan \textit{et al.} \cite{Wan}.

The systematic shift of the transition frequency is governed by the Zeeman shift (\textless 450(10)~kHz) and the ac Stark shift of the P$_{1/2}$ level due to the repump lasers (45(5)~kHz) while systematic shifts arising from the line shape and experimental controls are below 10 kHz. Utilizing only linearly polarized laser light for cooling and re-pumping, Zeeman splitting should only contribute to the line broadening rather than the systematic level shift. Thus, we expect the line shift to be significantly smaller than the Zeeman splitting of (\textless 450(10)~kHz). To determine the systematic shift due to the experiment control, we carefully measured the timings of the electronic pulses and determined the response of the laser modulators. Due to the short response times and comparable long probe times, these effects are estimated to be below 10~kHz. For shorter probe times the contribution of these shifts to the error budget increases. Therefore, the optimal probe time is a compromise between systematic shifts due to the experiment control and the shift of the line due to heating induced asymmetry of the line shape. Furthermore, the total interrogation time to obtain a sufficient signal-to-noise ratio and the intrinsic stability of the laser are also important factors to determine the optimal settings. In our system, a probe time of 25~$\mu$s at a repetition rate of 5~kHz has been found to be optimal.

\section{\label{sec:conclusion}Conclusion}

We have developed a scheme to measure the full fluorescence spectrum of dipole allowed transitions in trapped ions without the detrimental effects of laser induced heating. Utilizing this scheme, transition parameters such as linewidth and transition frequency can be determined more precisely than previously possible. By employing a differential probing of the fluorescence spectrum, a dispersion-like shape can be obtained which is well suited to determining the line center of the transition and can serve as an error signal for the stabilization of a laser. It also mitigates the effect of laser intensity fluctuations.
Employing the differential technique, we have stabilized a laser to the atomic transition and measured its frequency stability with an independent ion-trap system. We obtained a frequency instability of better than 10$^{-10}$ for an interrogation time of 1000~s.


\noindent\textbf{Acknowledgements} This work was supported by the Engineering and Physical Sciences Research Council (EPSRC) of the UK.

%
%

\end{document}